\documentclass[twocolumn]{aastex701}
\usepackage{graphicx,amsmath,amssymb}
\usepackage{natbib}
\usepackage{hyperref}
\hypersetup{breaklinks=true}  % set the option after loading
\usepackage{mathrsfs}
\usepackage{xcolor}
\usepackage{bm}
\usepackage{ulem}

\newcommand{\msun}{M_{\odot}}

\def\mnras{Mon. Not. R. Astron. Soc.}
\def\apj{Astrophys. J.}
\def\apjl{Astrophys. J. Lett.}

\def\aap{Astron. Astrophys.}

\shorttitle{Superdiffusion at the Galactic Centre}
\shortauthors{Amaro Seoane et al.}

\begin{document}

\title{Superdiffusion at the Galactic Centre}

\author[orcid=0000-0003-3993-3249,gname='Pau',sname='Amaro Seoane']{Pau Amaro Seoane}
\affiliation{Universitat Polit{\`e}cnica de Val{\`e}ncia, C/Vera s/n, Val{\`e}ncia, 46022, Spain}
\affiliation{Max Planck Institute for Extraterrestrial Physics, Giessenbachstra{\ss}e 1, 85748 Garching, Germany}
\affiliation{Kavli Institute for Astronomy and Astrophysics at Peking University, Beijing, China}
\email[show]{amaro@upv.es}

\author[orcid=0000-0003-0949-7191,gname='Kostas',sname='Tzanavaris']{Kostas Tzanavaris}
\affiliation{Max Planck Institute for Gravitational Physics (Albert Einstein Institute), D-30167 Hannover, Germany}
\affiliation{Leibniz Universit{\"a}t Hannover, D-30167 Hannover, Germany}
\email[]{konstantinos.tzanavaris@aei.mpg.de}

\author[gname='Reinhard',sname='Genzel']{Reinhard Genzel}
\affiliation{Max Planck Institute for Extraterrestrial Physics, Giessenbachstra{\ss}e 1, 85748 Garching, Germany}
\affiliation{Departments of Physics \& Astronomy, Le Conte Hall, University of California, Berkeley, CA 94720, USA}
\email[]{genzel@mpe.mpg.de}

\author[gname='Frank',sname='Eisenhauer']{Frank Eisenhauer}
\affiliation{Max Planck Institute for Extraterrestrial Physics, Giessenbachstra{\ss}e 1, 85748 Garching, Germany}
\affiliation{Department of Physics, TUM School of Natural Sciences, Technical University of Munich, 85748 Garching, Germany}
\email[]{eisenhau@mpe.mpg.de}

\author[orcid=0000-0003-1572-0396,gname='Thomas',sname='Ott']{Thomas Ott}
\affiliation{Max Planck Institute for Extraterrestrial Physics, Giessenbachstra{\ss}e 1, 85748 Garching, Germany}
\email[]{ott@mpe.mpg.de}

\author[gname='Stefan',sname='Gillessen']{Stefan Gillessen}
\affiliation{Max Planck Institute for Extraterrestrial Physics, Giessenbachstra{\ss}e 1, 85748 Garching, Germany}
\email[]{ste@mpe.mpg.de}

\author[gname='Guillaume',sname='Bourdarot']{Guillaume Bourdarot}
\affiliation{Max Planck Institute for Extraterrestrial Physics, Giessenbachstra{\ss}e 1, 85748 Garching, Germany}
\email[]{gbourda@mpe.mpg.de}

\author[gname='Diogo',sname='Ribeiro']{Diogo C. Ribeiro}
\affiliation{Max Planck Institute for Extraterrestrial Physics, Giessenbachstra{\ss}e 1, 85748 Garching, Germany}
\email[]{ribeiro@mpe.mpg.de}

\author[gname='Matteo',sname='Sadun-Bordoni']{Matteo Sadun Bordoni}
\affiliation{Max Planck Institute for Extraterrestrial Physics, Giessenbachstra{\ss}e 1, 85748 Garching, Germany}
\email[]{msadun@mpe.mpg.de}

\author[gname='Simran',sname='Joharle']{Simran Joharle}
\affiliation{Max Planck Institute for Extraterrestrial Physics, Giessenbachstra{\ss}e 1, 85748 Garching, Germany}
\email[]{sjoharle@mpe.mpg.de}

\author[gname='Felix',sname='Mang']{Felix Mang}
\affiliation{Max Planck Institute for Extraterrestrial Physics, Giessenbachstra{\ss}e 1, 85748 Garching, Germany}
\affiliation{Department of Physics, TUM School of Natural Sciences, Technical University of Munich, 85748 Garching, Germany}
\email[]{fmang@mpe.mpg.de}

\author[gname='Andreas',sname='Burkert']{Andreas Burkert}
\affiliation{University Observatory, Faculty of Physics, Ludwig-Maximilians-Universit{\"a}t, Scheinerstra{\ss}e 1, 81679 Munich, Germany}
\affiliation{Max Planck Institute for Extraterrestrial Physics, Giessenbachstra{\ss}e 1, 85748 Garching, Germany}
\email[]{burkert@usm.lmu.de}

\author[gname='Jorge',sname='Cuadra']{Jorge Cuadra}
\affiliation{Universidad Adolfo Ib{\'a}\~nez, Av. Padre Hurtado 750, Vi\~na del Mar, Chile}
\affiliation{Millennium Nucleus on Transversal Research and Technology to Explore Supermassive Black Holes (TITANS), Chile}
\email[]{jorge.cuadra@uai.cl}

\author[orcid=0000-0002-9019-9951,gname=Diego, sname='Calderón']{Diego Calderón}
\affiliation{Max-Planck-Institut für Astrophysik, Karl-Schwarzschild-Straße 1, 85748 Garching, Germany}
\email[]{calderon@mpa-Garching.mpg.de}

\author[gname='Hagai',sname='Perets']{Hagai B. Perets}
\affiliation{Physics department, Technion - Israel Institute of Technology, Technion city, Haifa 3200002, Israel}
\email[]{hperets@technion.ac.il}

\author[gname='Tsvi',sname='Piran']{Tsvi Piran}
\affiliation{Racah Institute of Physics, The Hebrew University, Jerusalem 91904, Israel}
\email[]{tsvi.piran@mail.huji.ac.il}

\author[gname="Re'em",sname='Sari']{Re'em Sari}
\affiliation{Racah Institute of Physics, The Hebrew University, Jerusalem 91904, Israel}
\email[]{sari@mail.huji.ac.il}

\begin{abstract}
Tracking S-star cluster orbits around Sgr A* calibrates orbital transport models for space-borne gravitational wave detectors. Standard kinetic theories model this cluster via local Fokker-Planck equations, which predict that general relativistic precession halts angular momentum diffusion at the Schwarzschild barrier. Because inverse-square gravitational encounters generate a Holtsmark torque distribution with infinite variance, resonant relaxation operates as a space-fractional process governed by non-local L\'{e}vy flights. We simulate this superdiffusive continuous-time random walk using a Markov chain initialized with empirical S-star orbits, including the recently observd S301. Integro-differential fractional operators allow trajectories to cross regions of quenched local diffusion without density buildup at the barrier. Non-equilibrium regimes yield immediate linear flux growth, while secular tidal heating at periastron inflates stellar radii to shift disruption boundaries. Regularized backward integration of the fractional transport equation traces current phase space configurations back to initial deposition states, matching the energy requirements of the \emph{Fermi} bubbles. Relativistic precession does not suppress mass-ratio inspiral rates, which provides a model for event topologies in target galactic nuclei.
\end{abstract}


\section{Introduction}

The Galactic Centre hosts Sgr A*, a central black hole with a mass of $M_{\bullet} = 4.3 \times 10^6 M_\odot$, surrounded by the nuclear star cluster. This environment constitutes the only galactic nucleus where current instruments resolve and track individual stellar orbits. The S-star cluster \citep{GenzelEtAl10,GillessenEtAl2017}, including S301 \citep{S3012026}, provides a spatilaly resolved map of strong-field orbital transport. Operating as test particles, these stars trace the central potential and map the non-local statistical mechanics of the near-Keplerian phase space.

Continuous local diffusion formalisms typically model stellar orbits. At sub-parsec scales, where Sgr A* dominates the potential, coherent secular torques drive resonant relaxation and dictate long-term angular momentum transport. Standard analytical approaches map this evolution using local Fokker-Planck operators, which invoke the central limit theorem to assume that background potential fluctuations generate Gaussian processes with finite variance \citep{AmaroSeoane2025_FracDyn}.

This local assumption creates a theoretical contradiction when applied to the Galactic Centre. The inverse-square geometry of Newtonian gravity invalidates the central limit theorem. The fluctuating stellar background generates torques that conform to a Holtsmark probability distribution, where the power-law tail produces infinite variance \citep{Holtsmark1919,AmaroSeoane2025_FrDym}. Resonant relaxation \citep{RT96,HopmanAlexander06,FouvryBarOr2018,BarOrFouvry2018, FouvryEtAl2019_b,FouvryEtAl2019, FouvryEtAl2019_c} operates as a continuous-time random walk governed by the generalized central limit theorem. This physical constraint dictates that angular momentum evolution across the S-star cluster proceeds via Lévy flights or ``jumps'', a superdiffusive mechanism in which localized orbital variations alternate with non-local jumps across phase space \citep{AmaroSeoane2025_FracDyn}.

Replacing local diffusion with fractional kinetics addresses structural discrepancies in galactic dynamics. Local Fokker-Planck models predict a transport limit at high dimensionless potentials, termed ``the Schwarzschild barrier'', with the effect that rapid relativistic apsidal precession averages out background torques and halts orbital decay. The non-local geometry of L\'{e}vy flights enables stellar trajectories throughout the nuclear cluster to cross regions of quenched local diffusion, which sustains a continuous flux into high-eccentricity states \citep{AmaroSeoane2025_FracDyn}. This jump mechanism determines the rate at which S-stars undergo tidal disruption or decouple from the stochastic background to become sources of gravitational radiation in the extremes-mass regime \citep{Amaro-SeoaneLRR,AmaroSeoane2022,BabakEtAl2017,Amaro-SeoaneEtAl07}.

The orbit of the recently discovered S-star S301 provides a test particle for these non-local transport regimes within the broader S-star ensemble. With an apparent magnitude of $m_K = 19.3$, S301 occupies an orbit with a period of $8.7$ years and a periastron velocity of $25000\,$km/s, or $8\%$ of the speed of light \citep{S3012026}. This trajectory subjects the star to tidal forces that drive thermomechanical heat dissipation. This process inflates the stellar envelope and establishes a dynamic tidal disruption boundary for extended bodies in the cluster.

The orbital kinematics of S301 and its neighboring stars subject their trajectories to general relativistic corrections beyond leading-order precession. While measurements isolate corrections of order $(v/c)^2$, the orbital parameters of S301 increase its sensitivity to corrections of order $(v/c)^3$. This sensitivity maps the Kerr metric and frame-dragging effects induced by the spin of Sgr A*. 

The Galactic Centre is a reference galaxy for LISA \citep{Amaro-SeoaneEtAl2017}. Mapping the phase space transport of this resolved inner cluster with interferometers such as GRAVITY allows us to study transport topologies and detection rates.

\section{Fractional dynamics at the Galactic Center}

The nuclear star cluster provides a high-density dynamical environment where conventional approximations of angular momentum transport fail. In the related literature, the cumulative effect of gravitational encounters is evaluated as a continuous, Markovian diffusion process via the local Fokker-Planck formalism \citep[see e.g.][and references therein]{BinneyTremaine08}. The foundation of this approach relies on the central limit theorem to converge the sum of independent stochastic variables into a Gaussian distribution. This assumption requires the variance of the individual interactions to remain finite.

The primary interaction in a stellar system is the inverse-square gravitational force, which produces a Holtsmark probability distribution for the background interactions \citep{Holtsmark1919}. The Fourier transform of this force distribution exhibits a non-analytic dependence on the wave vector $k$, structured as $\exp(-C n |k|^{3/2})$, where $n$ represents the stellar density and $C$ is a constant. This property dictates that the variance of gravitational torques diverges, invalidating the central limit theorem and demonstrating a structural limitation in classical local theories \citep{AmaroSeoane2025_FracDyn}. The application of local models to environments with relativistic apsidal precession predicts a cessation of angular momentum transport, defined as the Schwarzschild barrier \citep{MerrittEtAl11,BarOrAlexander2014,BarOrAlexander2016}. Under a \textit{local} Fokker-Planck equation, this region acts as a transport boundary, limiting the predicted rates of high mass-ratio inspirals and tidal disruption events; in reality, this barrier is merely an artifact of the local approximation rather than a feature of physical systems \citep[as seen in the numerical simulations of][]{BremEtAl2014,AmaroSeoane2025_FrDym}.

To explain these predictive limitations, we apply the generalized central limit theorem \citep{MetzlerKlafter2000}. When stochastic steps possess infinite variance, the sum of coherent torques converges to a L\'{e}vy stable distribution. The transport mechanism operates as a continuous-time random walk, known as a L\'{e}vy flight. This superdiffusive process intersperses short-range steps in angular momentum with long-range displacements. The step magnitude $\Delta J$ follows a probability density function governed by the power law,

\begin{equation} \label{eq:levy_tail}
P(\Delta J) \sim |\Delta J|^{-5/2}, 
\end{equation}

\noindent
where we calculate the exponent from the stability index $\alpha = 3/2$. We apply discrete jumps to the normalized angular momentum $J = \sqrt{1 - e^{2}}$, where $e$ is the orbital eccentricity, enforcing orbital closure via the reflecting boundary condition $J \le 1$.

To model this non-local transport, and following the work of \cite{AmaroSeoane2025_FrDym}, we utilize the space-fractional Fokker-Planck equation \citep{BarkaiEtAl2000}. This kinetic equation replaces local spatial derivatives with the Riesz-Weyl fractional operator, yielding the relation,

\begin{equation} \label{eq:space_ffpe}
\frac{\partial f(J, t)}{\partial t} = \mathcal{L}_{\text{drift}} f + D(J) \mathbb{R}^{3/2} f(J, t), 
\end{equation}

\noindent
where $D(J)$ is the state-dependent diffusion coefficient and $\mathbb{R}^{3/2}$ is the symmetric fractional derivative. The Riemann-Liouville integral definition of this operator computes the fractional rate of change by integrating over the entire domain of the distribution function. This integro-differential structure maps the heavy-tailed jump distribution, allowing stellar orbits to bypass localized regions of quenched diffusion in single intgration steps. This non-local formulation maintains the flux of objects into high-eccentricity trajectories, establishing the theoretical framework to predict transport topology and event rates of EMRIs and TDEs.

The local Fokker-Planck equation models angular momentum diffusion as a Markovian process constrained by the central limit theorem. The propagator for this differential equation is a Gaussian distribution, restricting phase space transport to \textit{sequential, incremental variations} and causing an exponential attenuation of probability density at large step lengths.

Eq.~(\ref{eq:space_ffpe}) replaces local derivatives with the Riesz-Weyl fractional operator. This integro-differential structure generates a L\'{e}vy stable distribution for the transition probabilities. The distinction between these transport mechanisms emerges in the asymptotic behavior of their propagators. The fractional operator produces a power-law tail that scales as the inverse of the step magnitude, maintaining finite probability densities for large angular momentum variations.

\begin{figure}[ht]
\centering
\includegraphics[width=\columnwidth]{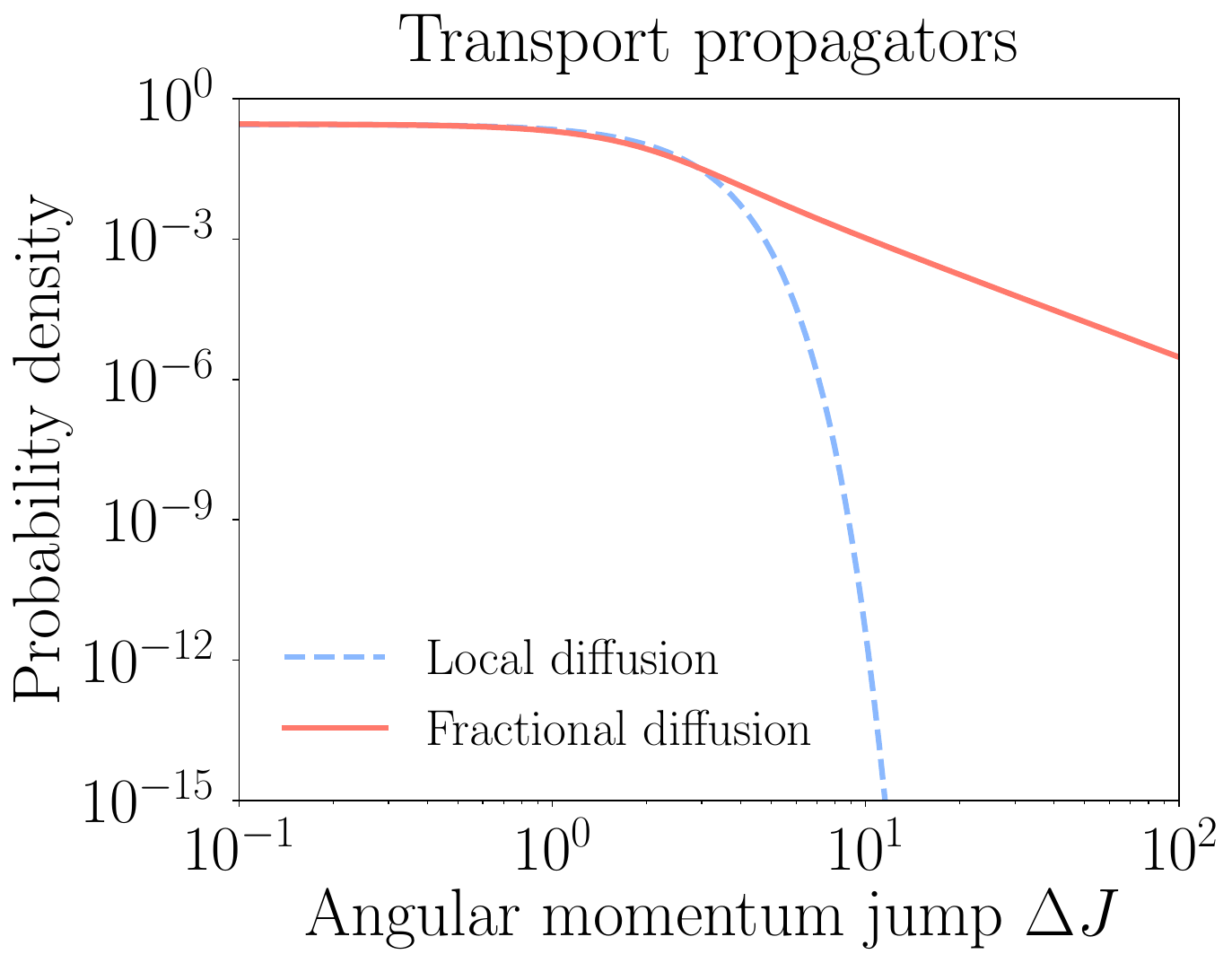}
\caption{Comparison of the propagators for local diffusion and fractional diffusion. We employ a Gaussian propagator corresponding to a diffusion coefficient $D=1$ and time $t=1$ for the local model. We utilize a L\'{e}vy stable distribution with a stability index $\alpha=3/2$ for the fractional model. We depict the probability densities on a single logarithmic scale to isolate the heavy-tailed asymptotic behavior. At a step magnitude $\Delta J = 10$, the local diffusion probability density attenuates exponentially to $10^{-12}$, while the fractional diffusion probability density follows a power law and measures $10^{-3}$, illustrating the superdiffusive transport capacity of the heavy tail.}
\label{fig:propagator_comparison}
\end{figure}

\noindent
Figure~\ref{fig:propagator_comparison} displays the transition probability distributions for local and fractional diffusion models. The local diffusion model confines particles to the core of the distribution, as the Gaussian tail decays exponentially. The fractional model maintains a heavy tail, generating non-local transport across the coordinate space. This structural difference provides the mechanism for stellar orbits to bypass regions of quenched local diffusion, such as the Schwarzschild barrier, in single integration steps.

\section{Evolution of the nuclear star cluster}

To evaluate the macroscopic transport topology of the Galactic Centre, we evolve the observed S-star ensemble over a simulated epoch of $30$ million years. We determine the statistical likelihood of individual objects intersecting the dynamic tidal disruption limits or completing gravitational wave inspirals. Because direct integration of dynamical systems over cosmological timescales exceeds computational capacities, we implement a discrete-time Markov chain to sample the space-fractional Fokker-Planck kinetics.

We initialize the ensemble using the empirical orbital parameters of $42$ S-stars, including S2 and S301 \citep{GillessenEtAl2017,S3012026}. We define the phase space coordinates using the dimensionless potential at periastron, $R_{\text{s}}/q$, and the fractional pericenter velocity, $\beta = v/c$, where $R_{\text{s}}$ represents the Schwarzschild radius of Sgr A* and $q = a(1-e)$ denotes the pericenter distance. 

We discretize the continuous-time random walk into temporal steps $\Delta t$. At each integration step, we subject the normalized angular momentum $J = \sqrt{1 - e^2}$ of every active particle to a stochastic displacement drawn from a L\'{e}vy stable probability distribution. We set the stability index to the Holtsmark value of $\alpha = 3/2$. The scale parameter $\sigma$ of the L\'{e}vy distribution reflects the superdiffusive scaling, proportional to the resonant relaxation diffusion coefficient $D_{\text{RR}}$ and the fractional time step,

\begin{equation} \label{eq:levy_scale}
\sigma = D_{\text{RR}} \Delta t^{2/3}\;.
\end{equation}

\noindent
We apply a reflecting boundary condition to map angular momentum values exceeding $1$ back into the physical domain, ensuring orbital closure.

Simultaneously, we calculate the deterministic secular evolution of the orbits. We employ the coupled differential equations to evaluate the continuous decay of the semi-major axis $a$ and eccentricity $e$ due to gravitational wave emission. We compute the energy and angular momentum dissipation rate $da/dt$, $de/dt$ as presented in the work of \cite{Peters64}.
We integrate these secular drifts across the timestep and superpose them onto the stochastic L\'{e}vy displacements. We additionally incorporate thermomechanical feedback from tidal forces acting on extended main sequence bodies \citep{PealeEtAl1979}. During highly eccentric orbits, repeated periastron passages deposit frictional energy into the stellar envelope. We evaluate the number of passages per timestep, $\Delta t / P$, where $P$ is the orbital period. We compute a fractional heating energy $\Delta E_{\text{heat}}$ that scales inversely with the sixth power of the pericenter distance,

\begin{equation} \label{eq:tidal_heating}
\Delta E_{\text{heat}} \propto \frac{\Delta t}{P} \left( \frac{r_{\text{t}}}{q} \right)^6\;,
\end{equation}

\noindent
where $r_{\text{t}}$ is the unperturbed tidal disruption radius. We derive this scaling from the quadrupolar approximation of the tidal potential. A central black hole of mass $M_{\bullet}$ exerts a differential gravitational force across a star of radius $r_{\star}$ and mass $m$. At the pericenter distance $q$, the amplitude of the tidal force scales as $F_{\text{tid}} \propto G M_{\bullet} r_{\star} / q^3$. During the periastron passage, the star responds dynamically, raising a tidal bulge with a fractional displacement $\delta r / r_{\star} \propto (M_{\bullet}/m) (r_{\star}/q)^3$. The mechanical work done by the tidal force to raise this bulge scales with the square of the displacement amplitude. By integrating the dissipated kinetic energy over the duration of the passage, the energy deposited into the stellar interior per orbit scales as the square of the tidal amplitude, producing an energy transfer $\Delta E \propto (G m^2 / r_{\star}) (M_{\bullet}/m)^2 (r_{\star}/q)^6$. We susbtitute the unperturbed tidal disruption radius, defined by the balance of self-gravity and tidal shear as $r_{\text{t}} = r_{\star}(M_{\bullet}/m)^{1/3}$, to isolate the dimensionless coordinate. Normalizing by the stellar binding energy $E_{\text{bind}} \propto G m^2 / r_{\star}$ calculates the fractional energy per passage $\Delta E / E_{\text{bind}} \propto (r_{\text{t}}/q)^6$. The secular accumulation of this energy scales with the orbital frequency $1/P$. As an illustration, if an orbit shifts its pericenter distance to half its original value, the energy deposited into the stellar envelope per orbit increases by a factor of $64$.

We map the accumulated energy to the inflation factor by modeling the expansion of the stellar envelope. We assume that a fraction of the deposited tidal heat $\Delta E_{\text{heat}}$ is retained within the stellar structure, augmenting its internal energy. The virial theorem relates the total energy of the star $E_{\text{tot}}$ to its radius via $E_{\text{tot}} \propto -G m^2 / r_{\star}$. The accumulation of heat alters the total energy to $E_{\text{tot}}' = E_{\text{tot}} + \Delta E_{\text{heat}}$. Setting this perturbed total energy equal to the new virial configuration $-G m^2 / R_{\star}'$ establishes the relationship for the perturbed radius $R_{\star}'$. We define the linear inflation factor $I(t) = R_{\star}' / r_{\star}$, which dictates the exact algebraic relation,

\begin{equation} \label{eq:inflation_factor}
I(t) = \left( 1 - \frac{\Delta E_{\text{heat}}(t)}{|E_{\text{tot}}|} \right)^{-1} \;.
\end{equation}

\noindent
As $\Delta E_{\text{heat}}(t)$ approaches the initial binding energy, the denominator decreases, driving $I(t)$ upwards. We impose the upper bound $I(t) \le 5$ based on stellar structure constraints. Standard main sequence stars cannot maintain hydrostatic equilibrium when inflated beyond this limit \citep[see e.g.][]{KW94}. Once the accumulated thermal energy drives the envelope expansion to a factor of $5$, the internal pressure gradients fail to counteract self-gravity and the external tidal forces. The star undergoes spontaneous mass loss via Roche lobe overflow or complete tidal disruption prior to reaching the formal geometric limit.

This radial inflation establishes a dynamic tidal disruption boundary $r_{\text{t,dyn}}(t)$ which migrates outward in phase space as the star absorbs tidal energy,

\begin{equation} \label{eq:dynamic_rt}
r_{\text{t,dyn}}(t) = I(t) r_{\text{t}}\;.
\end{equation}

\noindent
If a stochastic jump drives the pericenter distance below this dynamic threshold, the star undergoes tidal disruption.

We track the main sequence lifetimes of the S-stars, assigning temporal limits that follow a uniform distribution between $5$ and $35$ million years. Secular stellar evolution provides an additional mechanism for radial expansion. As stars exhaust their core hydrogen and evolve off the main sequence, their envelopes expand, which lowers their gravitational binding energy \citep[see e.g.][]{KW94}. This evolutionary expansion increases the cross section for partial or complete tidal disruption, providing a secondary channel for mass deposition into the central black hole prior to collapse. Upon exhausting this lifespan, a main sequence star undergoes gravitational collapse into a $10 \msun$ stellar-mass black hole. This state transition removes the extended gaseous envelope, which nullifies the dynamic tidal disruption limits and permits the compact remnant to penetrate deeper into the gravitational potential.

\begin{figure*}[ht]
\centering
\includegraphics[width=\textwidth]{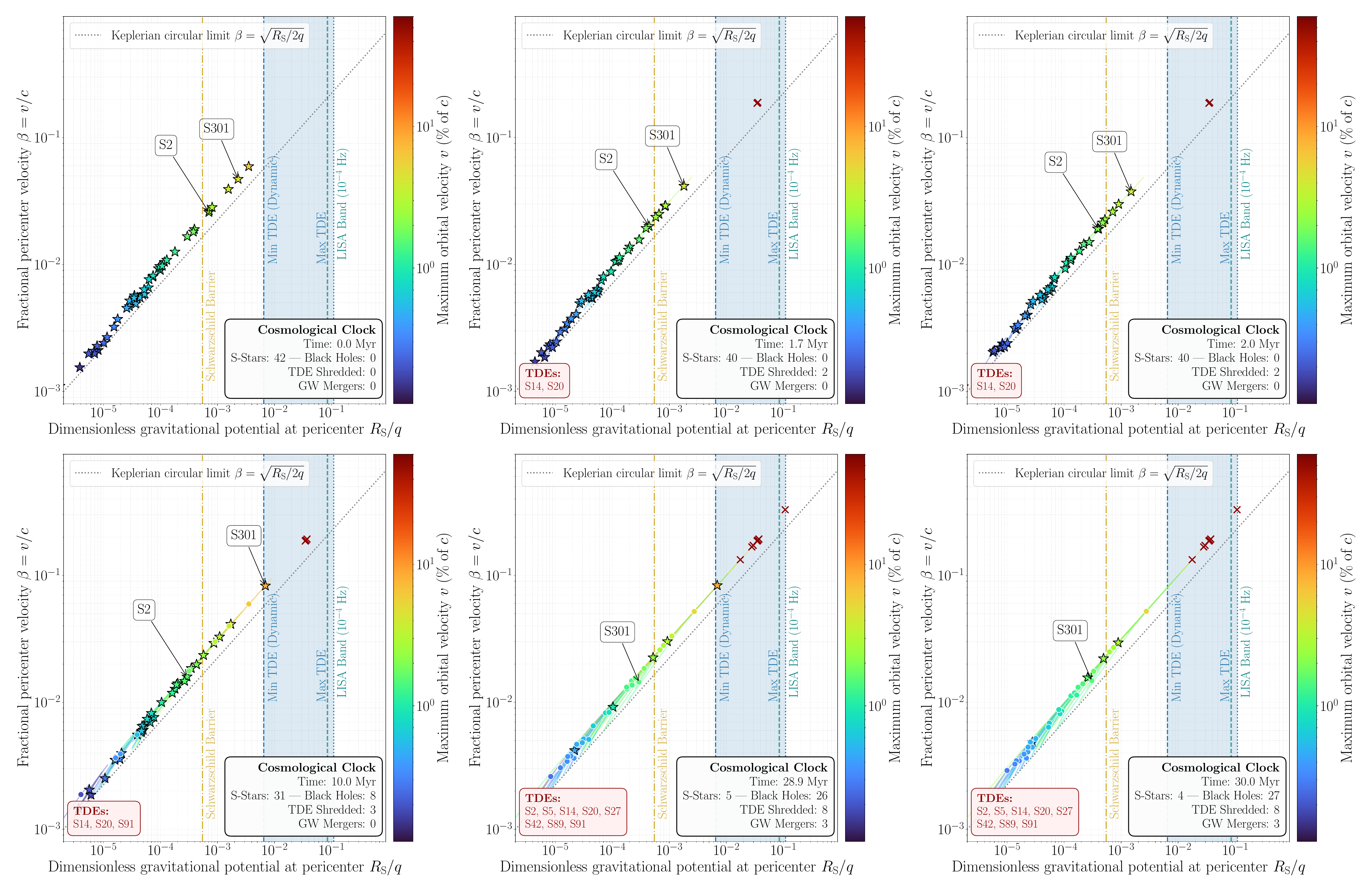}
\caption{Mosaic of six sequential snapshots detailing the phase space evolution of the S-star cluster. We display the dimensionless potential $R_{\text{s}}/q$ on the horizontal axis and the fractional pericenter velocity $\beta = v/c$ on the vertical axis. The gray dotted curve indicates the Keplerian circular velocity limit. The shaded blue region bounds the dynamic tidal disruption zone, limited by the minimum and maximum disruption potentials derived from the inflated stellar radii. The dash-dotted gold line marks the theoretical Schwarzschild ``barrier''. The dashed teal line specifies the entry frequency for LISA at $10^{-4}$ Hz. Star symbols denote main sequence stars, while circular markers with white edges represent stellar-mass black holes. The color map codes the fractional velocity $\beta$. Crosses indicate the exact phase space coordinates of tidal disruption events. A solid trajectory line traces the history of each particle. The boxes show instantaneous integration metrics, including the time and object counts. For an animation, see \url{https://youtu.be/l56mRpdMDDw} or \url{https://www.bilibili.com/video/BV1D9gk6aEKW/} .}
\label{fig:evolution_mosaic}
\end{figure*}

\noindent
Figure~\ref{fig:evolution_mosaic} visualizes this chronological progression across six sequential snapshots. In the initial phases, the main sequence stars execute L\'{e}vy flights across the angular momentum space. The heavy-tailed jump distribution produces non-local displacements that shift stellar orbits across the theoretical coordinate of the Schwarzschild barrier without localized accumulation,
meaning the buildup of phase space density directly adjacent to a transport barrier. In local Fokker-Planck formulations, the flux of particles $\mathcal{F}(J)$ relies on the local gradient of the distribution and a continuous diffusion coefficient $D(J)$,

\begin{equation} \label{eq:local_flux}
\mathcal{F}_{\text{local}}(J) = -D(J) \frac{\partial f(J, t)}{\partial J} \;.
\end{equation}

\noindent
At the Schwarzschild barrier, relativistic apsidal precession averages out the coherent background torques, driving the local diffusion coefficient $D(J)$ to zero. To maintain a non-zero steady-state flux across this barrier, Eq.~(\ref{eq:local_flux}) dictates that the gradient $\partial f / \partial J$ must diverge to infinity, forcing particles to pile up against the barrier coordinate.

Fractional dynamics eliminate this requirement. The non-local flux relies on the Riesz-Weyl operator, which integrates the distribution across the entire domain $s$,

\begin{equation} \label{eq:fractional_flux}
\mathcal{F}_{\text{frac}}(J) \propto \int_{J}^{1} \frac{f(s, t)}{(s - J)^{1/2}} \, ds \;.
\end{equation}

\noindent
Because this flux definition does not depend on the local derivative at $J_{\text{SB}}$, the distribution function does not need to steepen. A particle located at a higher angular momentum $s > J_{\text{SB}}$ executes a L\'{e}vy flight, bypassing the quenched diffusion zone entirely and landing directly at $J < J_{\text{SB}}$. This jump mechanism guarantees that phase space transport proceeds through the barrier, preventing any localized density accumulation at the Schwarzschild coordinate.

As the main sequence stars inflate due to thermomechanical feedback, their trajectories intersect the shifting dynamic limits, resulting in tidal disruption events. Stars that avoid tidal disruption complete their main sequence lifetimes and collapse into compact remnants. The compact radii of these black holes decouple them from the tidal disruption limits applicable to extended stars. We observe these remnants execute non-local transport deep into the potential field. At potentials exceeding the Schwarzschild barrier, the deterministic orbital decay from gravitational wave emission supersedes the magnitude of the stochastic resonant relaxation jumps. The remnants decouple from the continuous-time random walk and follow monotonic inspiral trajectories, crossing the $10^{-4}$ Hz frequency threshold into the detection band of space-borne interferometers, calculated by evaluating the characteristic frequency of the gravitational waves emitted during the inspiral phase. For a highly eccentric orbit, the gravitational wave emission spans a spectrum of harmonics of the orbital frequency, but the radiated power peaks at a frequency $f_{\text{GW}}$ corresponding to the timescale of the periastron passage. We compute this peak frequency using the mass of the central black hole $M_{\bullet}$ and the pericenter distance $q$,

\begin{equation} \label{eq:gw_frequency}
f_{\text{GW}} \approx \frac{1}{\pi} \sqrt{\frac{G M_{\bullet}}{q^3}} \;.
\end{equation}

\noindent
Space-borne interferometers, such as LISA, possess a design sensitivity window bounded at the lower end by acceleration noise. This noise floor restricts the detection of continuous gravitational wave signals to frequencies exceeding $10^{-4}$ Hz. Substituting the threshold frequency $f_{\text{GW}} = 10^{-4}$ Hz and the mass $M_{\bullet} = 4.3 \times 10^6 \msun$ into Eq.~(\ref{eq:gw_frequency}), we isolate the specific phase space boundary for the pericenter distance. When a stellar-mass black hole undergoes a non-local jump that reduces its pericenter distance below this critical coordinate, the energy dissipation rate driven by gravitational wave emission outpaces the characteristic timescale of resonant relaxation L\'{e}vy jumps. The orbit transitions from stochastic random walks to continuous, deterministic decay observable by space-borne interferometers.

\section{Backward integration and the active past of the Galactic Centre}

The Galactic Centre has very likely had a very energetic past. The \emph{Fermi} bubbles \citep{DoblerEtAl2010} require an energy release of approximately $10^{55}$ erg, potentially fueled by periodic star captures \citep{ChengEtAl2011} or active galactic nucleus jet activity $1$--$3$ million years ago \citep{GuoMathews2012}. Concurrently, young stellar rings imply the existence of a $10^4$ to $10^5 \msun$ self-gravitating accretion disk during the same epoch \citep{NayakshinCuadra2005}. We investigate whether the superdiffusive transport framework bridges these energetic requirements and the orbital evolution of the S-stars.

To fuel an outburst $1$ to $3$ million years ago, transport mechanisms must move mass to the central black hole on short timescales. Turbulent friction transports angular momentum outward and mass inward in standard thin accretion disks. The local viscous accretion timescale governs this process,

\begin{equation} \label{eq:viscous_timescale}
t_{\text{visc}} \sim (R/v_K)\alpha^{-1}(R/H)^2,
\end{equation}

\noindent
where $R$ represents the radial distance from the central black hole, $v_K$ denotes the local Keplerian velocity, $\alpha$ is the dimensionless viscosity parameter, and $H$ defines the vertical scale height of the disk. At a distance of $0.1$ parsec, this timescale ranges from $10^7$ to $10^9$ years, which exceeds the temporal limits for the transient.

Fractional superdiffusion offers an alternative. Stars born in the initial accretion disk interact through gravitational torques, experiencing non-local L\'{e}vy flights. These superdiffusive jumps rapidly drive the normalized angular momentum to low values, dropping stars directly into the tidal disruption zone. A sequence of such disruption events bypasses the local viscous accretion limit, injecting material into the central potential well on timescales compatible with the observed transients.

To test this mechanism, we trace the initial conditions of the S-star cluster backward in time. Standard local diffusion models cannot perform this tracing reliably. Gaussian differential operators smooth localized information, causing reverse integration to amplify high-frequency numerical artifacts. Under local transport assumptions, the structural memory of the injection event erases exponentially, rendering the backward problem ill-posed.

The theoretical framework of superdiffusion supports integrating the space-fractional Fokker-Planck equation backward in time with appropriate regularization. We clarify the validity of this backward integration by evaluating the behavior of the fractional propagator in the frequency domain. Applying a spatial Fourier transform to the symmetric space-fractional diffusion equation generates an ordinary differential equation for the characteristic function $\hat{f}(k, t)$ in terms of the wave vector $k$. Evolving forward from an initial state at $t=0$, the solution is

\begin{equation} \label{eq:fourier_forward}
\hat{f}(k, t) = \hat{f}(k, 0) \exp\left(-D |k|^\alpha t\right)\;.
\end{equation}

\noindent
The negative sign in the exponent of Eq.~(\ref{eq:fourier_forward}) dictates that high-frequency modes decay, which corresponds to the physical process of the distribution spreading out as time moves forward. To reconstruct a past state at a lookback time $\tau = -t > 0$, we invert the propagator algebraically, producing

\begin{equation} \label{eq:inverse_propagator}
\hat{f}(k, \tau) = \hat{f}(k, 0) \exp(D |k|^\alpha \tau) \;.
\end{equation}

\noindent
The positive sign in Eq.~(\ref{eq:inverse_propagator}) is mathematically required to reverse the diffusion process, as it amplifies the high-frequency modes to restore the sharpness of the original distribution. In standard local diffusion ($\alpha = 2$), the term $\exp(D k^2 \tau)$ acts as a divergent amplification factor for high wave numbers $k$, causing the inverse integral to become undefined for any distribution containing standard high-frequency numerical noise. This divergence renders backward Gaussian diffusion strictly ill-posed.

In fractional superdiffusion, the stability index $\alpha = 3/2$ reduces the asymptotic growth rate of the backward propagator to $\exp(D |k|^{3/2} \tau)$. This sub-quadratic scaling permits the application of a low-pass Gaussian regularization filter $W(k)$ to stabilize the inverse operator. The product $\exp(D |k|^{3/2} \tau) W(k)$ converges to zero as $k \to \infty$. This condition guarantees that the inverse Fourier transform produces a finite, integrable solution. We formulate the regularized backward density as

\begin{equation} \label{eq:backward_proof_solution}
f(J, \tau) = \frac{1}{2\pi} \int_{-\infty}^{\infty} \hat{f}(k, 0) e^{D |k|^{3/2} \tau} W(k) e^{-i k J} \, dk \;.
\end{equation}

\noindent
Because the integral converges absolutely, the backward mapping produces a mathematically stable phase space topology.

\begin{figure}[ht]
\centering
\includegraphics[width=\columnwidth]{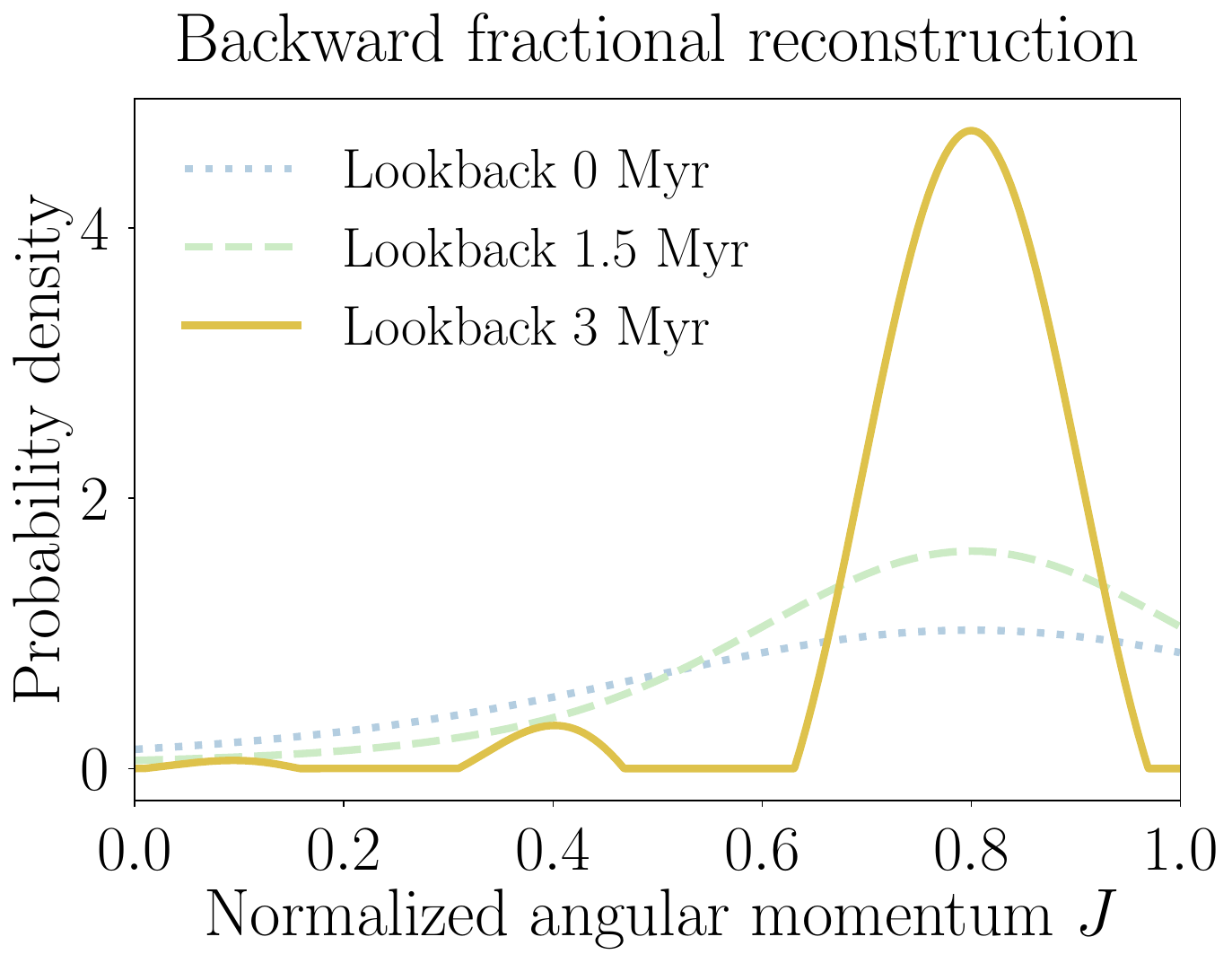}
\caption{Reconstruction of historical phase space topologies via regularized backward integration of the space-fractional Fokker-Planck equation. The horizontal axis measures the normalized angular momentum $J$, and the vertical axis denotes the probability density. The current state at lookback time $0$ Myr exhibits a broad, superdiffusive distribution across the phase space. Reversing the fractional transport operators to a lookback time of $1.5$ Myr narrows the distribution. At a lookback time of $3$ Myr, the fractional operator collapses the density into a localized peak, providing a mathematical reconstruction of a possible initial state for the stellar ensemble.}
\label{fig:backward_integration}
\end{figure}

\noindent
Figure~\ref{fig:backward_integration} depicts the quantitative results of this backward procedure. The numerical calculation directly evaluates Eq.~(\ref{eq:backward_proof_solution}) in Fourier space. The inverse integration takes a broad, nearly uniform angular momentum distribution (lookback time $0$ Myr) and evolves it backward. Rather than encountering uniform noise amplification, the fractional distribution sharpens. At a lookback time of $3$ million years, the distribution collapses into a localized peak near $J = 1/2$.

This reconstruction evaluates the fractional transport mechanism by maintaining variables such as time-dependent potentials, gas hydrodynamics, and non-isotropic relaxation fields constant. Furthermore, the framework applies to an ex situ origin for the S-stars, such as the Hills mechanism \citep{Hills88}, where the tidal disruption of binary systems deposits individual stars into orbits with high eccentricity. If binary captures produce the S-star cluster, superdiffusion operates on the post-capture population and modifies their initial angular momenta over time. Under this assumption, the localized peak in the reconstruction represents the initial orbital distribution of the captured stellar ensemble rather than an in situ accretion disk. The calculations from this fractional framework align with independent observational and theoretical constraints \citep{DoblerEtAl2010,ChengEtAl2011,GuoMathews2012,NayakshinCuadra2005}. The reconstructed timeline outlines a scenario where a localized stellar population, whether originating in a discrete disk or forming through binary captures, dissolves via superdiffusion and supplies stars into the loss cone to fuel the transients in the Galactic Centre.

\section{Conclusions}

The Galactic Centre provides the only physical environment where astronomical instruments track the phase space transport of individual stars in real time. This spatial resolution reveals a limitation in classical kinetic theories. When modelled via local Fokker-Planck equations, relativistic apsidal precession drives the local diffusion coefficient to zero at the Schwarszchild barrier, a result not observed in direct-summation $N$-body simulations \citep{BremEtAl2014,AmaroSeoane2025_FrDym}. This classical approximation halts the inward flux of stellar mass and stops the supply of compact objects to the central black hole. The space-fractional formulation removes this boundary. Because inverse-square gravitational encounters produce a Holtsmark torque distribution with infinite variance, resonant relaxation operates as a continuous-time random walk. Under the generalized central limit theorem, stellar trajectories execute non-local L\'{e}vy flights. These discrete phase space jumps cross the region of quenched local diffusion and maintain the transport flux through the relativistic barrier.

Within this non-local transport topology, high-eccentricity main sequence stars such as S301 serve as empirical probes of sub-parsec dynamics. As the trajectory of S301 penetrates the strong-field regime, the star undergoes high-magnitude energy dissipation during close periastron passages. Tracking the orbital evolution of such extended bodies demonstrates how thermomechanical feedback inflates the stellar envelope, establishing a dynamic rather than static tidal disruption boundary. The orbital elements of S301 provide an observable benchmark for an object that executes discrete spatial jumps, persisting in a high-shear tidal environment where classical continuous drift predicts destruction.

To map the exact topological transition between stochastic transport and deterministic general relativistic inspiral, observational astronomy must locate stars occupying deeper, higher-eccentricity orbits. Identifying stellar populations with pericenters directly adjacent to the Schwarzschild barrier supplies the test particles necessary to empirically evaluate the superdiffusive bypass mechanism. Characterizing these deep-pericenter trajectories isolates the specific phase space coordinates where L\'{e}vy flights transition into continuous orbital decay driven by gravitational wave emission. Tracking this high-eccentricity population maps the direct precursor pathways for high mass-ratio inspirals, supplying the observational data required to calibrate detection rate projections for space-borne interferometers such as LISA.

Applying these fractional transport operators backward in time transforms the orbital kinematics of the S-star cluster into a diagnostic record of past energetic events. Because the integro-differential Riesz-Weyl operators preserve the structural scaling of non-local transport, reversing the stochastic displacements maps the high-eccentricity orbits back to localized initial deposition states. This regularized backward integration establishes a mathematically stable method to reconstruct historical mass-deposition transients. By tracing the current phase space topology to its initial configuration, this framework provides a mechanism to calculate the timeline of the accretion episodes that supplied the energy required to inflate the \emph{Fermi} bubbles and generate historical X-ray flares.

\begin{acknowledgments}
PAS acknowledges support by China’s National Foreign Expert Program (H). DC and JC acknowledge the financial support from ANID-FONDECYT Regular 1251444. The research of DC has been funded by the Alexander von Humboldt Foundation.
\end{acknowledgments}

\end{document}